# Absence of orbital current torque in a magnetic metal/Pt/$CuO_x$ trilayer

Zhihao Yan[1,2], Lujun Zhu[3], Xiangrui Qin[1,2], Zhengxiao Li[1,2], Lijun Zhu[1,2*]

1. *State Key Laboratory of Semiconductor Physics and Chip Technologies, Institute of Semiconductors, Chinese Academy of Sciences, Beijing 100083, China*

2. *University of Chinese Academy of Sciences, Beijing 100049, China*

3. *College of Physics and Information Technology, Shaanxi Normal University, Xi'an 710062, China*

*ljzhu@semi.ac.cn

Naturally oxidized copper ($CuO_x$) was cited as an exceptionally strong orbital-current source despite the keen debate over the concept of orbital current. Here, we report unambiguous experimental evidence for the absence of a detectable orbital-current torque in the Fe/Pt/$CuO_x$ trilayers. Using a magnetic metal detector of Fe thin film with a negligible self-induced torque, the damping-like torque due to Pt/$CuO_x$ bilayers is clarified to entirely arise from the spin Hall spin current of the Pt. As the Pt thickness increases, the torque for Fe/Pt/$CuO_x$ and Fe/Pt/MgO increases coherently and monotonically, as exactly expected from the spin Hall effect of Pt and the drift-diffusion model of spin angular momentum. These findings suggest poor generality or even universal absence of orbital current torque.

The development of fast, low-power memory and computing has triggered considerable interest in the spin-orbit torques (SOTs) exerted on a nanomagnet by spin currents [1-10]. In the past decades, orbital polarization was consensually believed to be irrelevant to non-local angular momentum transport effects [11], in contrast to spin polarization that flows as a spin current. Recently, there is an increasingly blooming interest in challenging the fundamental question as to whether there was an "orbital current" that can somehow flow non-locally between adjacent materials [12-15]. However, there has also been many theories [16-19] and experiments [20-25] against the concept of orbital current, which suggest that orbital polarization, if generated by the bulk orbital Hall or interfacial gradient effects, should be strongly localized.

A prototype orbital current candidate is $CuO_x$ within magnet/Pt/$CuO_x$ trilayers, in which the $CuO_x$ was formed by natural oxidation of a 3 nm Cu [26-29]. In a previous hysteresis loop shift analysis of TmIG/Pt/$CuO_x$ trilayers [26], the Pt/$CuO_x$ interface was suggested to inject an orbital current into the magnetic oxide TmIG via the mediation of the strong-spin-orbit-coupling Pt layer (with thickness $t_{Pt}$) and to ultimately exert "orbital torque" on the TmIG. However, recent studies have revealed that the hysteresis loop shift includes multiple contributions *irrelevant to angular momentum transfer* (e.g., perpendicular Oersted field [30] or perpendicular DMI fields [31-33]). In addition, a non-monotonic $t_{Pt}$ dependence in the spin-pumping inverse spin Hall voltage and THz emission signals of the magnet/Pt/$CuO_x$ [28,29] is also a typical consequence of spin transport in the case of a constant spin Hall ratio [34-36]. Therefore, it has remained a critical open question as to the existence of orbital current in the magnet/Pt/$CuO_x$.

In this work, utilizing the metallic 3*d* Fe with negligible self-induced torque, we for the first time establish that the SOT of the Fe/Pt/$CuO_x$ increases monotonically with $t_{Pt}$, coinciding with that of Fe/Pt/MgO without $CuO_x$. These results provide robust evidence for the dominance of the spin Hall effect and the absence of orbital current torque in the Fe/Pt/$CuO_x$.

For this work, we sputter-deposited Fe (1.6 nm)/Pt (0-5.5 nm)/Cu (3 nm) trilayers (termed as Fe/Pt/$CuO_x$) and Fe (1.6 nm)/Pt (0-5.5 nm)/MgO (1.6 nm)/Ta (1.5 nm) trilayers (termed as Fe/Pt/MgO) on oxidized silicon wafers, with the top Cu and Ta layers oxidized after exposure to atmosphere. In Fig. 1b we show representative scanning transmission electron microscopy (STEM) images of the Fe/Pt/$CuO_x$ and the Fe/Pt/MgO, suggesting reasonably sharp interfaces. After the growth, the natural oxidation process of Cu was monitored by the increase of the resistance of a strip with time (Fig.1c). The oxidation of the Cu is essentially completed at 125 hours. The samples were fabricated into 5×60 $\mu m^2$ Hall bar devices. Figure 1d shows the energy dispersive X-ray spectroscopy (EDS) of the Fe/Pt/$CuO_x$ Hall device and a control substrate. From superconducting quantum interference device measurements, the Fe/Pt/$CuO_x$ and the Fe/Pt/MgO samples have in-plane magnetic anisotropy (Fig. 1e), while the saturation magnetization ($M_s$) is 1507±51 emu/$cm^3$ for the Fe/Pt/$CuO_x$ (Fig. 1f) and 1688±78 emu/$cm^3$ for the Fe/Pt/MgO. The 10% variation in the $M_s$ values should arise from the sample fabrication.

The SOTs of the samples were quantified from harmonic Hall voltage measurement [24,37]. Under the excitation of a sinusoidal electric field ($E$) of 33.3 kV/m (Fig. 2a), the first harmonic Hall voltage ($V_{1\omega}$) is first collected as a function of the out-of-plane field ($H_z$) (Fig. 2b) for the determination of the macrospin behavior, the anomalous Hall voltage ($V_{AHE}$), and the in-plane anisotropy field ($H_k$). As shown in Fig. 2c, $V_{AHE}$ decreases dramatically as the Pt thickness increases, while $H_k$ varies little. The $V_{AHE}$ and $H_k$ values also exhibit a small variation between the Fe/Pt/$CuO_x$ series and Fe/Pt/MgO series, which we mainly attribute to the sample fabrication. The partially oxidized $CuO_x$ may have residual conductance, leading to weaker interfacial scattering and reduced resistivity enhancement in the Pt layer than the fully insulating MgO.

The harmonic Hall voltage ($V_{2\omega}$) is then collected (Fig. 2d) as a function of the azimuth angle ($\varphi$) of the in-plane magnetic field ($H_{xy}$=1.5-3.5 kOe) relative to the current direction. $V_{2\omega}$ of a macrospin is given by [37]

$$V_{2\omega} = V_{DL+ANE}\cos\varphi + V_{FL+Oe}\cos\varphi\cos2\varphi + V_{PNE}\sin2\varphi, \quad (1)$$

$$\text{with } V_{DL+ANE} = V_{AHE}H_{DL}/2(H_{xy} - H_k) + V_{ANE}. \quad (2)$$

Here, $H_{DL}$ is the damping-like SOT field, $V_{ANE}$ is the anomalous Nernst voltage induced by the vertical thermal gradient, $V_{PNE}$ is the planar Nernst voltage induced by the longitudinal thermal gradient. The fit of the $\varphi$ dependence of the $V_{2\omega}$ data to Eq. (1) yields the values of $V_{DL+ANE}$ and $V_{FL+Oe}$ for each magnitude of $H_{xy}$. As shown in Fig. 2e, $H_{DL}$ is determined from the slope of the linear dependence of $V_{\mathrm{DL+ANE}}$ vs $V_{\mathrm{AHE}}/2(H_{xy} - H_{\mathrm{k}})$ following Eq. (2). With the values of $H_{\mathrm{DL}}$, the damping-like SOT efficiency per electric field, $\xi_{\mathrm{DL}}^{E}$, is calculated following [38]

$$\xi_{\mathrm{DL}}^{E} = (2e/\hbar)\mu_0 M_s t H_{\mathrm{DL}}/E, \qquad (3)$$

where $e$ is the elementary charge, $\hbar$ the reduced Planck constant, $\mu_0$ the permeability of vacuum, and $t$ the thickness of the magnetic layer.

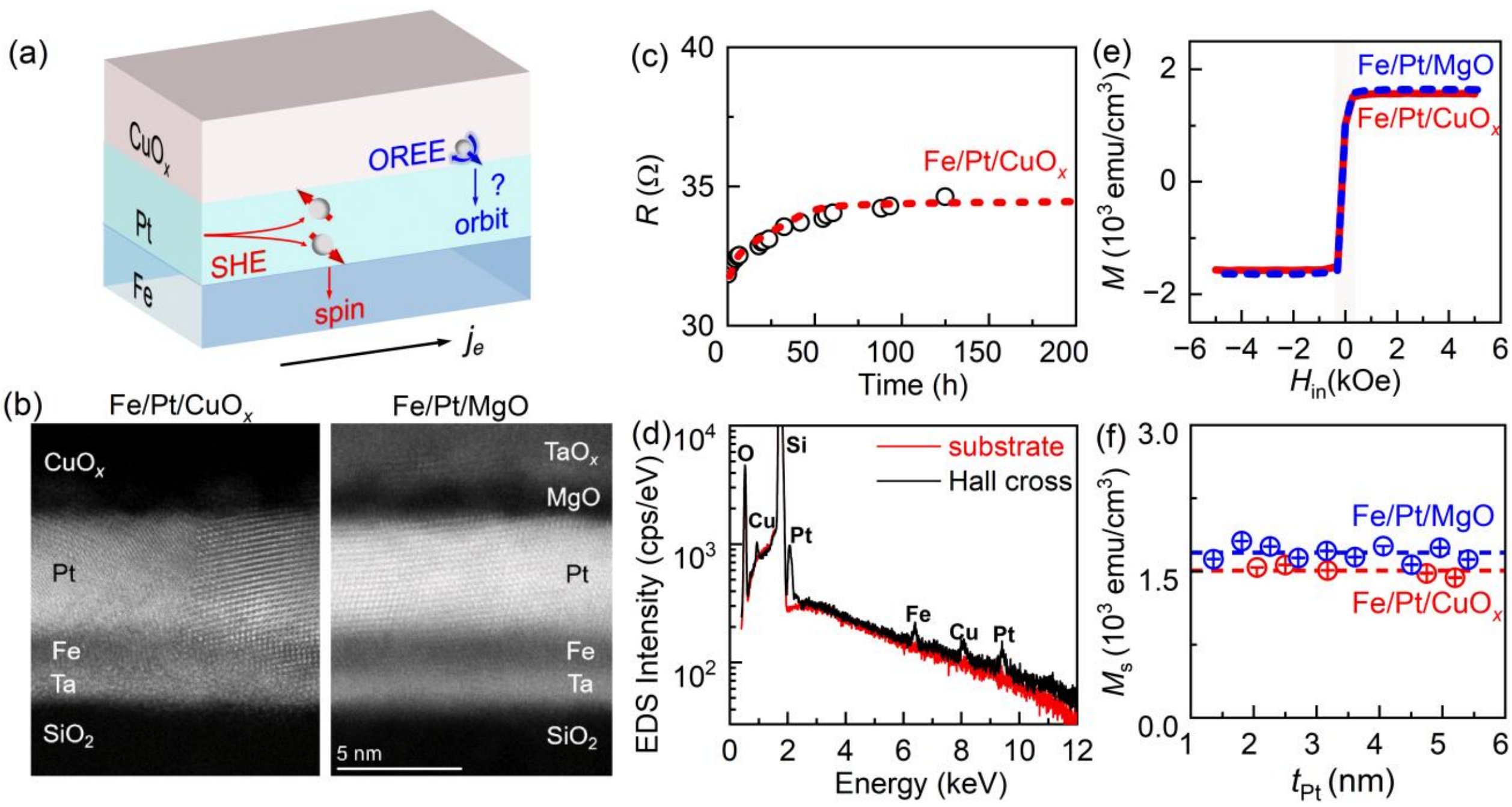

**Fig. 1.** (a) Potential spin and orbital transport in the Fe/Pt/CuO$_x$. (b) STEM images of the Fe/Pt (4.5nm)/CuO$_x$ and Fe/Pt (4.5nm)/MgO. (c) Resistance of a Fe/Pt (1.5nm)/CuO$_x$ film vs time. (d) EDS of the Fe/Pt/CuO$_x$ Hall bar. (e) Magnetization vs the in-plane magnetic field. (f) Saturation magnetization vs the Pt thickness.

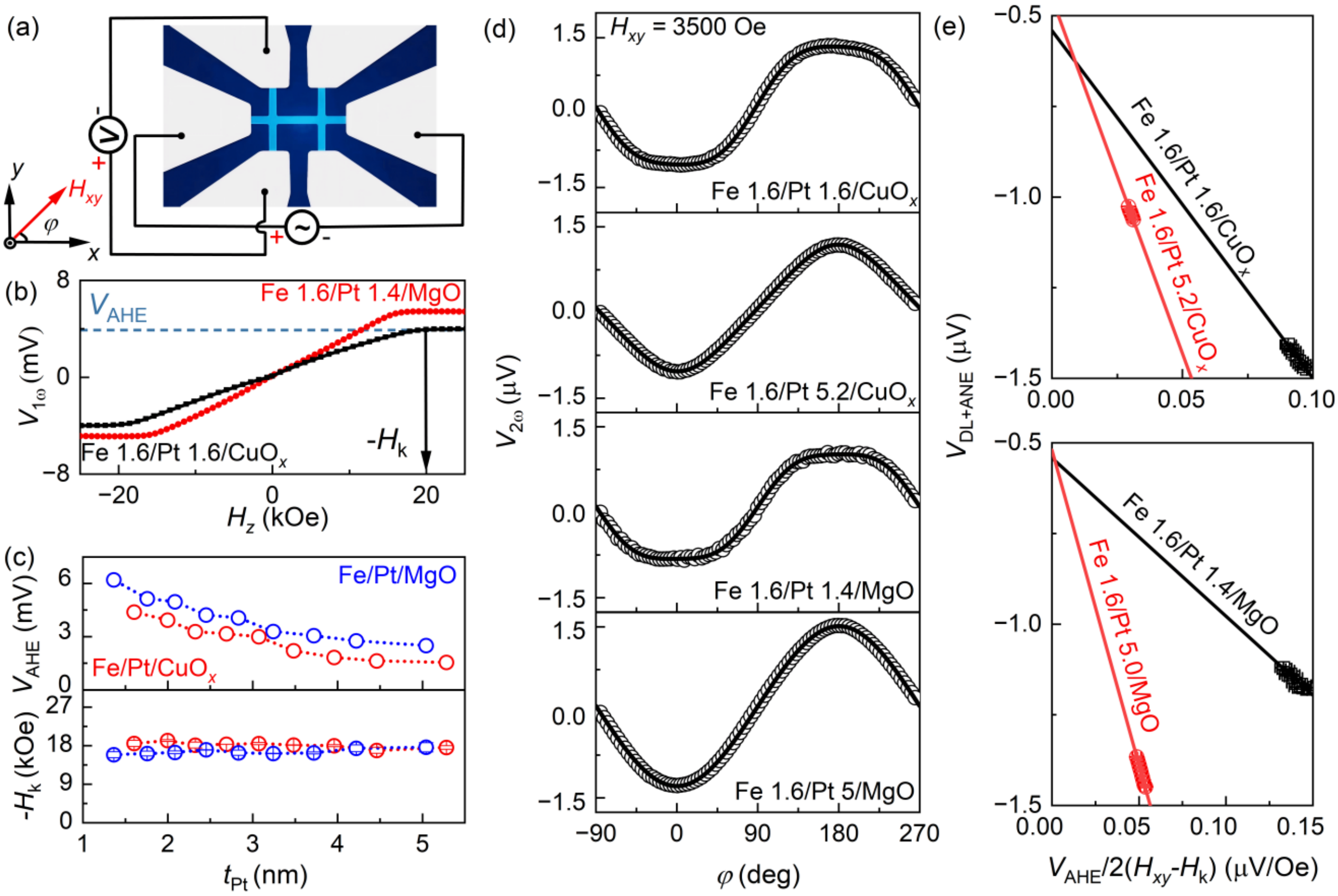

**Fig. 2.** (a) Optical microscopy image of the Hall bar device and the measurement coordinate. (b) First harmonic Hall voltage vs perpendicular magnetic field. (c) $V_{\mathrm{AHE}}$ and $H_{\mathrm{k}}$. (d) $\varphi$ Second harmonic Hall voltage vs in-plane magnetic field angle. (d) $V_{\mathrm{DL+ANE}}$ vs $V_{\mathrm{AHE}}/2(H_{xy}$-$H_{\mathrm{k}})$.

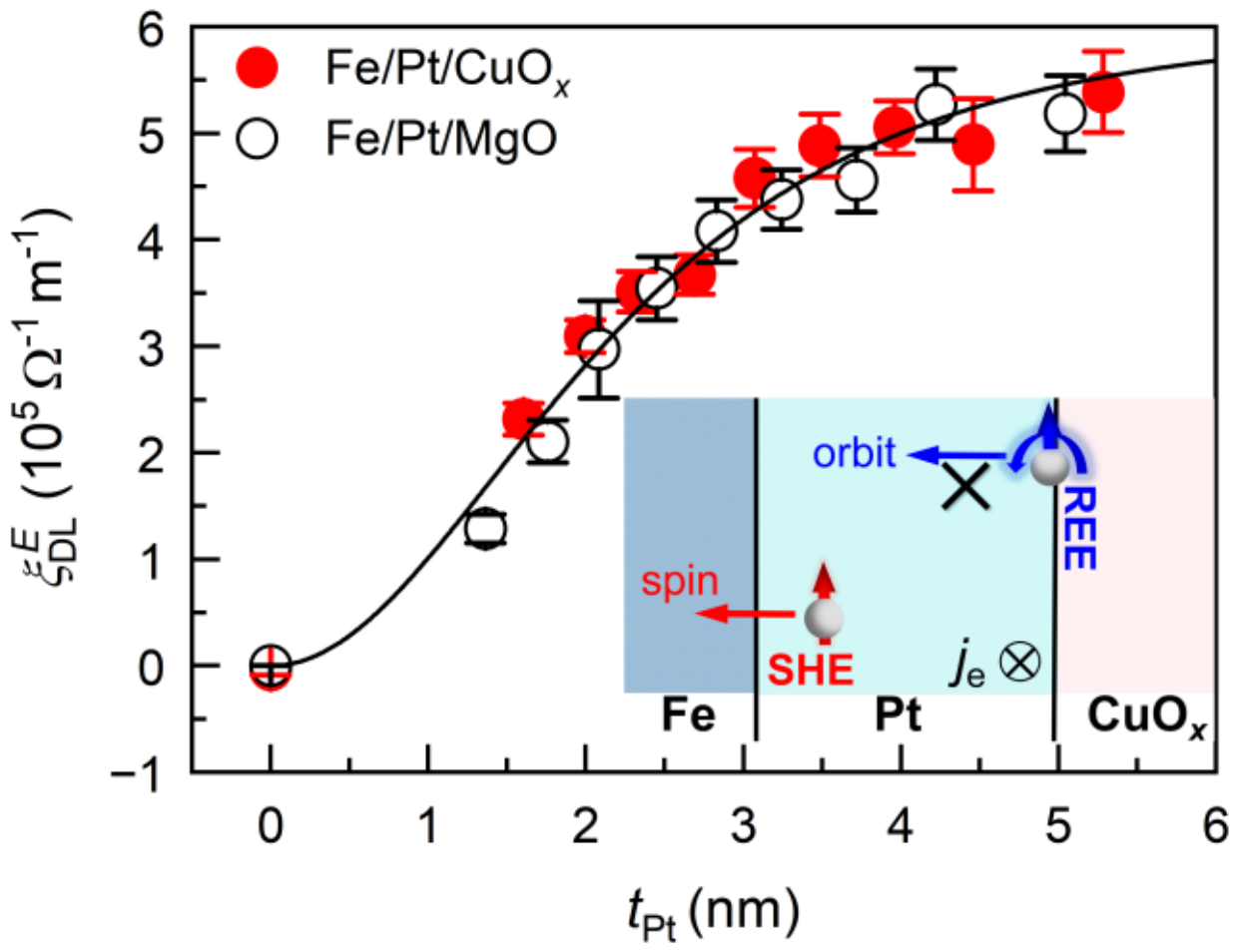

**Fig. 3.** Spin-orbit torque efficiency plotted as a function of the thickness of Pt for Fe/Pt/CuO$_x$ and Fe/Pt/MgO. Inset: the spin Hall spin current flows from Pt into Fe, while there is no orbital flow from the Pt/CuO$_x$ interface.

As shown in the Fig. 3, $\xi^E_{DL}$ for both the Fe/Pt/CuO$_x$ and the Fe/Pt/MgO increases monotonically from essentially zero at $t_{Pt}$ = 0 nm towards $5.4\times10^5$ $\Omega^{-1}m^{-1}$ as $t_{Pt}$ further increase. The absence of any significant torque for Fe/CuO$_x$ without a Pt layer agrees with the negligible torque in various FM/MgO devices [39]. The monotonic increase of $\xi^E_{DL}$ with $t_{Pt}$ agrees well with the drift-diffusion model of spin Hall spin current and can be fit by [3]

$$\xi^E_{DL} = \sigma_{SH}\,[1\text{-sech}(t_{Pt}/\lambda_s)](1+\tanh(t_{Pt}/\lambda_s)/2\lambda_s\rho_{xx}G^{\uparrow\downarrow})^{-1}. \quad (4)$$

With the average Pt resistivity of $\rho_{xx}\approx 40.0$ μΩ cm, the bare spin-mixing conductance of $G^{\uparrow\downarrow}\approx 5.9\times10^{14}$ $\Omega^{-1}m^{-2}$ for the Fe/Pt interface [39,40], the best fit of the thickness dependent $\xi^E_{DL}$ data yields a diffusion length of $\lambda_s\approx$ 1.8±0.1 nm and Hall conductivity of $\sigma_{SH}$ = $(1.36\pm0.04)\times10^6$ $\Omega^{-1}$ m$^{-1}$ for angular momentum related to the SOT of the Fe/Pt/CuO$_x$ and the Fe/Pt/MgO, which are consistent with the previous report of the spin diffusion length and spin Hall conductivity of Pt [3,10,41].

These results establish that only the spin current from the spin Hall effect of Pt participated in the generation of the SOT in the Fe/Pt/CuO$_x$ and the Fe/Pt/MgO and that there is no indication of any orbital current torque. As schematically illustrated in Fig. 3, the absence of the orbital current torque most likely suggests strong localization of orbital polarization, if any, or even absence of orbital generation at the Pt/CuO$_x$ interface.

Finally, we note that our conclusion of negligible orbital current is influenced by the different thickness scales of the unusual magnetoresistance of $Ni_{81}Fe_{19}$/Pt and $Ni_{81}Fe_{19}$/CuO$_x$ (without a strong SOC layer for effective orbital-spin conversion) bilayers in a previous experiment [42], since unusual magnetoresistance is universally expected from electron scattering at *magnetic interfaces* of trilayer, bilayers, and single layers in the absence of angular momentum transport [43]. The spin torque-ferromagnetic resonance signals of $Ni_{81}Fe_{19}$/CuO$_x$ [44] may be potentially from the self-induced torque of $Ni_{81}Fe_{19}$ that was recently found significant [20].

In summary, we have demonstrated unambiguous evidence for the absence of orbital current torque due to the CuO$_x$ and its interface. High-quality experiments reveal that the dampinglike SOT of the Fe/Pt/CuO$_x$ trilayers arises predominantly from the spin Hall spin current of the Pt. The torque increases with the Pt thickness as expected from the spin drift-diffusion model, whether or not there was a CuO$_x$ layer. These results have advanced the understanding of the orbital transport in the CuO$_x$ systems that were considered as the prototype orbital material. The consistent absence of the orbital current torque in the FM/Pt/CuO$_x$ systems and previously reported Ta [20], Al [21], and Ti [23-25] is highly suggestive of the poor generality or even universal absence of orbital current. This work will stimulate precise verification of the existence of orbital current in other systems.

The authors acknowledge Kai Chang and Yizheng Wu for helpful suggestions. This work was supported by the Beijing Natural Science Foundation (Z230006) and the National Natural Science Foundation of China (12274405).